\documentclass[aps,prb,preprint, amsmath,amssymp,showpacs]{revtex4-1}

\usepackage[T1]{fontenc}
\usepackage[latin1]{inputenc}
\usepackage[pdftex]{graphicx,color}
\usepackage{rotating}
\usepackage{caption}
\usepackage{bm}       
\usepackage{amssymb}   
\usepackage{amsmath}
\usepackage{bm}
\usepackage{CJK}
\usepackage{times}
\def\k1{k_1}
\def\k2{k_2}
\def\q1{q_1}
\def\q2{q_2}

\def\({\left (}
\def\){\right )}
\def\[{\left [}
\def\]{\right ]}

\newcommand{\beq}{\begin{equation}}
\newcommand{\eeq}{\end{equation}}

\begin{document}
\date{\today}
\flushbottom \draft
\title{Excitation of Biomolecules with Incoherent Light: Quantum yield for
the photoisomerization of model retinal}

\author{T. V. Tscherbul and P. Brumer}
\affiliation{Chemical Physics Theory Group, Department of Chemistry, and Center for Quantum Information and Quantum Control, University of Toronto, Toronto, Ontario, M5S 3H6, Canada}\email[]{ttscherb@chem.utoronto.ca}

\begin{abstract}
\textit{Cis-trans} isomerization in retinal, the first step in vision,
is often computationally studied 
from a time dependent viewpoint. Motivation for such studies
lies in coherent pulsed laser experiments that explore
the isomerization dynamics. However, such biological processes
take place naturally in the presence of incoherent light, which
excites a non-evolving mixture of stationary states. 					
Here the isomerization problem is considered from the latter
viewpoint and applied to a standard two-state, two-mode linear vibronic coupling
model of retinal that explicitly includes a conical intersection between the ground and first excited electronic states. The
calculated quantum yield at 500 nm agrees well with both the previous time-dependent calculations of Hahn and Stock (0.63) and with experiment ($0.65\pm0.01$), as does its wavelength dependence.  
Significantly, the effects of environmental relaxation on the quantum yield
in this well-established model are found to be negligible.
The results make clear the connection of the
photoisomerization quantum yield to properties of
stationary eigenstates, providing alternate insights into conditions for yield
optimization.

\end{abstract}

\maketitle
\clearpage
\newpage

\vspace{1cm}
{\bf TOC graphic}

\begin{figure}[b]
	\centering
	\includegraphics[width=0.8\textwidth, trim = 0 0 0 0]{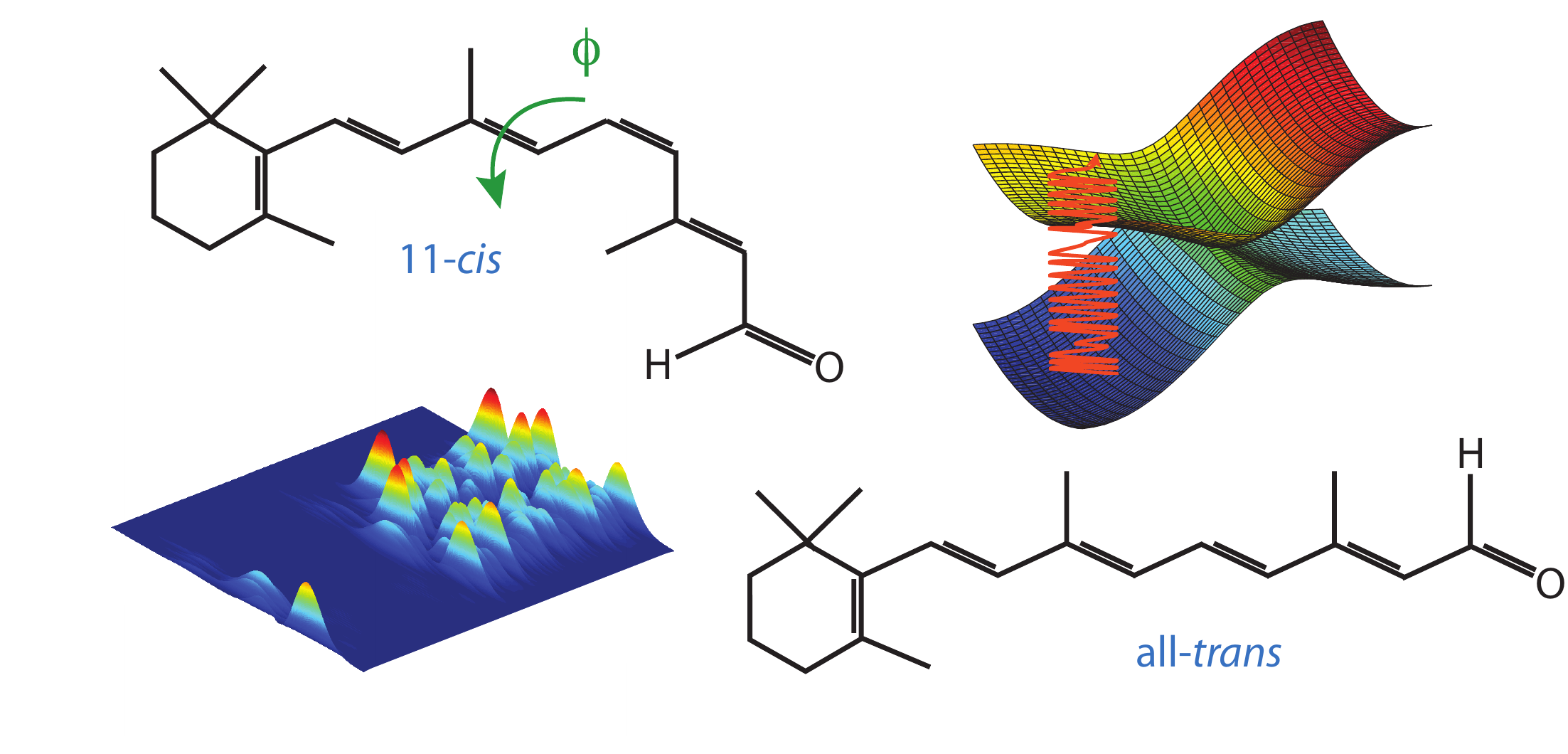}
	\caption*{}
\end{figure}

\newpage

\section{Introduction}
\label{Introduction}

Light-induced 11-{\it cis} $\to$ all-{\it trans} isomerization
in retinal is a paradigmatic example of an important ultrafast photochemical
reaction in biology
\cite{MathiesReview,Mathies94,Mathies10,Stock00,Stock00a,Stock03,Stock04,Stock05}. The photoreaction is the first step in
dim-light vision, and its
 high quantum yield, and
formation of all-{\it trans} product within 200 fs, contribute to
the high efficiency of the phototransduction cascade
\cite{MathiesReview,Mathies94,Mathies10,Stock00,Stock03,Stock05}.
The importance of retinal photoisomerization has
made its mechanism the subject of numerous experimental and
computational studies. A related biological process, {\it cis-trans} isomerization
in Retinoic Acid, has become of increasing interest in its role in
zebrafish hindbrain development\cite{Bensimon12}.

To study retinal dynamics,  current pump-probe experiments use ultrashort femtosecond 
laser pulses to excite the retinal from the ground
($S_0)$ to the first excited ($S_1$) electronic state. The subsequent
isomerization dynamics is then followed using a series of probe pulses,
providing important insights into the sub-200 fs timescale of photoproduct
formation \cite{Mathies94,Mathies10}, the coherent dynamics of
all-\textit{trans} photoproduct \cite{Mathies94}, and most recently,
the role of conical intersections in determining the reaction rate \cite{Mathies10}.
Significantly, retinal photoisomerization
occurs with  high quantum yield ($Y=0.65\pm 0.01$), making
the first step in visual phototransduction highly efficient
\cite{Baylor}. Measurements of the quantum yield $Y(\lambda)$
as a function of excitation wavelength $\lambda$ show
a maximum at 500 nm and a small decline of $\sim$5\% as the laser
wavelength increases from 500 nm to 570 nm \cite{Mathies00,Mathies03}.

A detailed time-dependent computational study of retinal
photoisomerization using a multilevel Redfield theory
\cite{Stock00,Stock03,Stock05} based on a minimal two-state,
two-mode (2D) model of retinal chromophore\cite{Stock00} provided
insights into the role of avoided crossings, conical
intersections, and dissipative dynamics in ultrafast energy
conversion in biomolecules \cite{Stock03}. In addition, they
qualitatively reproduce many salient features of
isomerization dynamics of retinal in rhodopsin, including the
high quantum yield. There and
elsewhere the quantum yield is defined dynamically, as the probability of
forming the all-{\it trans} product starting from the initial {\it
cis} wavepacket created by an ultrafast Franck-Condon excitation from the
ground {\it cis} state.





This time-dependent wavepacket view  \cite{MathiesReview,Mathies00,Mathies03} is
well-suited to describe time-domain experiments on rhodopsin,
in which ultrafast $S_0 \to S_1$ excitation creates a wavepacket on a highly
excited state.
However, in nature,
the excitation involves natural light. Such light has a very short
coherence time (1.32 fs for sunlight) compared to the fs
laser pulses used experimentally \cite{Mathies94,Mathies10},
and is incident on the molecule for a far longer time. As a result, 
after some time, which is dependent on a number of conditions\cite{Zaheen,cmb,Grinev}, the molecule
is prepared in a mixture of stationary states that
exhibits no coherent time evolution\cite{JiangBrumer,BrumerShapiro}.
Furthermore, these stationary states are, from a zeroth-order viewpoint,
a linear superposition of {\it cis} and {\it trans} configurations.
The existence and some unusual properties (such as microsecond lifetimes) of 
stationary eigenstates of this kind  in large polyatomic molecules
were examined experimentally as early as 1977 \cite{ZewailPNAS,ZewailSL}.

The questions then arise as to how to extract and understand the
quantum yield in the case of incoherent light excitation,
and how to describe the \textit{cis-trans} isomerization of retinal from a stationary eigenstate viewpoint that
is appropriate for vision under natural light conditions. A substantial step towards
this major goal is provided in this paper.





Here, we address this issue by first generalizing
the concept of the quantum yield to the case of incoherent light excitation of a polyatomic molecule.
We note that in the limit of rapid decoherence of initial
coherences in the reduced density matrix, the
photoreaction quantum yield is stationary. This
definition is then applied to calculate the quantum yield of {\it cis-trans}
photoisomerization of retinal using a minimal two-state,
two-mode model\cite{Stock00}. We find
excellent overall agreement between our calculated quantum yield and the previous time-dependent result of Stock and Hahn \cite{Stock00}.
 In addition, our calculations, in both the presence and absence of relaxation,
 agree with the observed value of the quantum yield at 500 nm and qualitatively
reproduce its observed decline with
increasing excitation wavelength \cite{Mathies00}. These
findings demonstrate the role of localization properties of the stationary
states in determining the quantum yields, a crucial feature in
the process induced with natural light. 

Note that there are various definitions of the 
quantum yield that are adopted in time-dependent studies. Below, 
we consistently utilize the parameters and approach of Ref. \citenum{Stock00},
which has become something of a ``standard model'' for basic retinal dynamics.
This imposes a number of consistency  requirements that are discussed in
Sect. \ref{summary}.

Before proceeding further, we emphasize that incoherent excitation of biomolecules
embedded in condensed-phase environments involves two processes that occur simultaneously: (a) the creation of the stationary states by the incoherent light, and (b) the relaxation between the stationary states caused by the interaction with the environment.  In this work, we treat these processes sequentially, assuming that the stationary states are formed first and then relax due to the system-environment coupling. This allows for a better assessment of their  individual roles. The eigenstates formed on stage (a) are considered in Secs. II and III; the effects of relaxation are considered in Sec. IV and, significantly, shown to have a negligible effect on the  quantum yield of retinal photoisomerization treated within in the two-state, two-mode model \cite{Stock00}.
The results, as seen below, motivate a future study of simultaneous incoherent excitation
and relaxation in this and related model retinal systems.

 \section{The Quantum Yield}
 \label{quantumyield}

We extract the quantum yield for the stationary case from the
standard time dependent result, hence exposing their relationship.
Reference \citenum{Stock00} defines the photoreaction quantum yield from the time-dependent view 
as 
\begin{equation}\label{Y}
Y = \frac{P^{(1)}_\text{trans}(t) }{ P^{(0)}_\text{cis}(t) + P^{(1)}_\text{trans}(t) } \quad (t\to \infty),
\end{equation}
where the time-dependent populations of 11-{\it cis} and all-{\it trans} isomers are defined as \cite{Stock00,Stock03,Stock04}
\begin{equation}\label{populations}
P^{(n)}_\alpha(t) = \text{Tr} \{ \rho(t) \hat{P}^{(n)}_\alpha\}\,.
\end{equation}
Here $\rho(t)$ is the reduced density matrix of the retinal subsystem, $\alpha=$ 11-{\it cis} or $\alpha=$ all-{\it trans}, and
\begin{align}\label{projectors}\notag
\hat{P}^{(0)}_\text{cis} = \Theta(\pi/2 - |\phi|) | \psi_0 \rangle \langle \psi_0| \\
\hat{P}^{(1)}_\text{trans} = \Theta( |\phi| - \pi/2) | \psi_1 \rangle \langle \psi_1|\,.
\end{align}
are the projection operators, which divide the full range of
the isomerization coordinate $\phi\in [-\pi/2, 3\pi/2]$ into
the {\it cis} ($\phi \in [-\pi/2,\pi/2]$) and {\it trans} regions ($\phi \in [\pi/2,3\pi/2]$), as illustrated in
Fig. 1. In Eq. (\ref{populations}), $\Theta(x)$ is
the Heaviside step function, and the operators $| \psi_n
\rangle \langle \psi_n|$ project onto the ground and excited
diabatic electronic
states $|\psi_{0}\rangle$ and $|\psi_{1}\rangle$. Qualitatively, $P^{(0)}_\text{cis}$
can be thought of as projecting onto the adiabatic ground state reactant \textit{cis} region and 
$P^{(1)}_\text{trans}$ onto the adiabatic ground state product \textit{trans}.
In this definition, the adiabatic excited state populations  
$P^{(1)}_\text{cis}$ and $P^{(0)}_\text{trans}$ are neglected assuming that these populations have decayed to zero in the long-time limit. 
Alternative definitions [e.g., including $P^{(1)}_\text{cis}$ in the numerator and
$P^{(0)}_\text{trans}$ in the denominator of Eq. (\ref{Y})] are certainly possible and will be explored in future work \cite{Tscherbul},
and would necessitate a refitting of the potential parameters to the new definition
of the time-dependent quantum yield.



As an example, consider impulsive FC excitation of
retinal from its ground electronic and vibrational states to
the first excited electronic state $|\psi_1\rangle$. Here the 
density matrix of the system at time zero is given by
\cite{Stock00,Stock03,Stock05}
\begin{equation}\label{rho_t0}
\rho(t=0) = |\psi_1\rangle |00\rangle \langle 00| \langle \psi_1|
\end{equation}
where $|00\rangle = |n_\phi = 0, n_x = 0\rangle$ is the ground state
with zero quanta in the torsional and coupling modes $n_\phi$ and $n_x$.

It is convenient to express the  density matrix in the
system eigenstate basis defined by
$\hat{H}_S |i\rangle = \epsilon_i |i\rangle$
where $\epsilon_i$ and $|i\rangle$ are time-independent
eigenvalues and eigenfunctions. In this basis, with
frequencies $\omega_{ij} = (\epsilon_i - \epsilon_j)/\hbar$,
$\rho_{ij}(t)= \rho_{ij}(0) e^{-i \omega_{ij} t}$.
Hence the population dynamics is given by
\begin{equation}\label{P1t_trans}
P^{(n)}_\alpha(t) = \sum_{i,\,j} e^{-i \omega_{ij}t} \langle i | \rho(0) | j\rangle \langle j |\hat{P}^{(n)}_\alpha |i\rangle 
\end{equation}

Since the time-dependence arises from the oscillatory behavior
of the coherences, (i.e., terms with $i\ne j$), if
$\langle i|\rho(0)|j\rangle = 0$ then the state populations are time independent.
An important example of such a case is molecular excitation with
incoherent light [Eq.  (\ref{P1t_trans})],
which populates, after some time, the eigenstates of $\hat{H}_S$ without
any subsequent coherent dynamics \cite{JiangBrumer}. Similarly, environmentally induced effects
cause the loss of coherences.  In this case,
retaining only the diagonal terms in Eq. (\ref{P1t_trans}), we have
\begin{equation}\label{Pdiag}
P^{(n)}_\alpha = \sum_{i} \langle i | \rho(0) | i\rangle \langle i |\hat{P}^{(n)}_\alpha |i\rangle
\end{equation}

This expression may be made more physically transparent by
noting that the diagonal elements of the reduced density matrix
following impulsive FC excitation at $t=0$ correspond to the
linear absorption spectrum of the molecule described by the
spectral lineshape function \cite{Zare} $A(\omega_i)$, where $\omega_i=(\epsilon_i-\epsilon_0)/
\hbar$ is the excitation  frequency measured  from  the  ground
vibrational  state  of  the {\it cis}-isomer with energy $\epsilon_0$ (we assume that this is the only vibrational state populated prior to excitation).
 Then  it follows from Eq. (\ref{rho_t0}) that
\begin{equation}\label{rho_me}
\langle i|\rho(0) |i\rangle = \langle i |\psi_1\rangle |00\rangle \langle 00| \langle \psi_1 |i\rangle = |\langle i| \hat{\mu} |\psi_0\rangle |00\rangle|^2 =  A(\omega_i)
\end{equation}
where
 $\hat{\mu} = \mu_{10} | \psi_1\rangle \langle \psi_0|$ is the transition dipole moment operator, and we set $\mu_{10} = 1$ a.u. without loss of generality [since the $|\mu_{10}|^2$ prefactors multiplying state populations cancel out in the expression for $Y$ in Eq. (\ref{Y})].  Therefore, Eq. (\ref{Pdiag}) may be rewritten 
 in the form
\begin{equation}\label{PI}
P^{(n)}_\alpha = \sum_{i} A(\omega_i) \langle i |\hat{P}^{(n)}_\alpha |i\rangle
\end{equation}
Hence, as expected, only the eigenvalues $\epsilon_i$ that have
non-negligible probability of being excited by incoherent light 
make a contribution to the stationary populations in Eq.
(\ref{PI}).

Given Eq. (\ref{PI}), the time-dependent definition of the
quantum yield (\ref{Y}) extends to the frequency domain by defining
\begin{equation}\label{QY1}
Y_1= \frac{P^{(1)}_\text{trans} }{ P^{(0)}_\text{cis} + P^{(1)}_\text{trans} }
\end{equation}
with the ``pre-averaged'' populations $P^{(n)}_\alpha$ given by
Eq. (\ref{PI}). Alternatively, we can first
define   a   frequency-dependent   {\it   trans}  /  {\it  cis}
probability ratio
\begin{equation}\label{QY2omega}
Y(\omega_i) = \frac{P^{(1)}_\text{trans}(\omega_i) }{ P^{(0)}_\text{cis} (\omega_i) + P^{(1)}_\text{trans}(\omega_i) },
\end{equation}
where $P^{(n)}_\alpha(\omega_i) = \langle i | \hat{P}^{(n)}_\alpha|i\rangle$ are the expectation values of the projection operators in Eq. (\ref{PI}). The resulting frequency-dependent quantum yield (\ref{QY2omega}) can be averaged with the normalized spectral line shape function to give
\begin{equation}\label{QY2}
Y_2 = \sum_i A(\omega_i) Y(\omega_i)
\end{equation}
It is clear that the definitions (\ref{QY1}) and (\ref{QY2}),
termed the  ``pre-averaged" and ``post-averaged" quantum yield,
are not equivalent, but both provide a physically meaningful
measure of (frequency-dependent) photoreaction efficiency. 
Specifically, the pre-averaged definition (\ref{QY1}) is consistent with the original 
definition of Hahn and Stock [Eq. (\ref{Y})] \cite{Stock00}, whereas the post-averaged definition admits a clear physical interpretation as the degree of {\it cis} vs. {\it trans} character of a collection of eigenstates independently excited by incoherent light. Expressions 
related to Eqs. (\ref{QY1}) and (\ref{QY2}) were previously obtained 
in References \citenum{MK} and \citenum{MS},  where 
the long-time limits of electronic state populations were associated with the
phase volumes occupied by the wavepackets evolving on the upper and lower 
adiabatic potential energy surfaces.


Equations (\ref{P1t_trans}), (\ref{PI}), (\ref{QY1}), and
(\ref{QY2}) form the central tool for the computations below.
They generalize the quantum yield in Eq. (\ref{Y}) to the
frequency domain relevant to stationary eigenstates, and provide a theoretical basis for the
computational study of photoinduced isomerization in model retinal as
described below. The essential physics comes from
the recognition that the established time dependent definition
[Eq. (\ref{Y})] relies entirely on the projections onto
domains of $\phi$ to define \textit{cis} vs. \textit{trans}
configurations.	 Indeed, each of Eqs. (\ref{Y}), (\ref{QY1}) and (\ref{QY2}) 
have essentially the same meaning, the fraction of population in the 
ground adiabatic state that is, in the long time limit, in the {\it trans}-configuration, disregarding any initial population in ground-state {\it cis} well.
This is achieved in Eq. (\ref{Y}) by placing all population in an initial wavepacket
on the excited electronic surface. Equations (\ref{QY1}) and (\ref{QY2omega}) achieve this by putting
all population initially into a stationary mixture on the excited potential
surface.
What the latter two equations emphasize is that
in the presence of incoherent light and decoherence, the appropriate states to consider
in the long time limit are the stationary eigenstates of the Hamiltonian. They, as seen below, will allow stationary state
insight into the efficiency of the isomerization process.

 Below, we apply this formulation to evaluate
the quantum yield for the primary photoreaction in rhodopsin
using the two-state two-mode model\cite{Domcke}, which was previously applied \cite{Stock00}
to  linear absorption \cite{Stock00a}, Raman
\cite{Stock00a}, and femtosecond pump-probe \cite{Stock04}
spectra of retinal in rhodopsin.
The  two-state two-mode model and its
multidimensional (25-mode) extension \cite{Stock00a}
have since been used to explore quantum
dynamics and coherent control of {\it cis-trans}
photoisomerization of retinal chromophore in rhodopsin
\cite{Abe,Batista,Arango}. 
We note that the model has the following advantages over 
the one-dimensional scenario (e.g. Ref. \citenum{HokiBrumer}):
(1) the present model is two-dimensional, including a relevant bend degree of freedom; 
(2) it accounts for the conical intersection between the ground and the first excited electronic states of retinal; (3) it reproduces many salient features of the experimentally measured isomerization dynamics, including the ultrafast 200 fs timescale, the transient pump-probe spectra, and the energy storage of the photoreaction.

\section{Computational Results}

We emphasize that the two-state two-mode model 
and the associated definition of the quantum yield [Eq. (\ref{Y})] were 
previously parametrized so as to reproduce various properties of the photoreaction\cite{Stock00a}
of retinal in rhodopsin. It neglects the effects of solvation, which are known to dramatically affect the photoisomerization 
dynamics in solution \cite{Turro,Ruhman,Hynes}, but does include, due to the parameter fit,
some features of the interaction of the two modes with the remaining molecular background. Our study is thus restricted 
to {\it cis-trans} photoisomerisation of retinal in rhodopsin.

The   wavefunctions  $|i\rangle $  in  Eq.
(\ref{PI})  are  the eigenfunctions of the 2D model Hamiltonian
given by \cite{Stock00,Stock03,Domcke}
\begin{equation}\label{model}
H_S = T\delta_{nn'} +
\begin{pmatrix}
E_0 + \frac{1}{2}\tilde{V}_0(1-\cos\phi) + \frac{\omega}{2}x^2 & \lambda x \\
\lambda x & E_1 - \frac{1}{2}\tilde{V}_1(1-\cos\phi) + \frac{\omega}{2}x^2 + \kappa x
\end{pmatrix}.
\end{equation}
where $T=-\frac{1}{2m}\frac{\partial^2}{\partial \phi^2} + \frac{\omega}{2}\frac{\partial^2}{\partial x^2}$ is the kinetic energy operator,  $\phi$  is  the tuning  mode  (or  generalized reaction coordinate) corresponding to low-frequency torsional modes, and
$x$  is  the  coupling  mode that corresponds to high-frequency
unreactive  modes.  The second term on the right-hand side of Eq. (\ref{model}) is the interaction potential in a basis spanned by the diabatic electronic functions $|\psi_n\rangle$ with $n=0,1$ \cite{Stock00}.  A plot  of  the adiabatic PESs obtained by
diagonalizing the potential energy term in Eq. (\ref{model}) is shown in Fig. 1.
The  model parameters $E_n$, $\tilde{V}_n$, $\omega$, $\kappa$,
and  $\lambda$,  chosen to reproduce the femtosecond
dynamics of retinal in rhodopsin \cite{Stock00,Stock03}, are (in eV): $E_0=0$, $E_1=2.48$, $\tilde{V}_0=3.6$, $\tilde{V}_1=1.09$, $\omega=0.19$, $\kappa=0.1$,  $\lambda=0.19$, and $m^{-1}=4.84\times 10^{-4}$. 

A total of 900 eigenvalues and eigenvectors of $\hat{H}_S$ were
calculated  and  used  to  assemble  the matrix elements of the
projector   operators   and  the  lineshape  function  in  Eqs.
(\ref{QY1})   and   (\ref{QY2}).   The  converged
$A(\omega)$  was  in agreement with results in Ref. \citenum{Stock00}.
Interestingly, we found that the spectrum could be classified as
integrable, insofar as the nearest neighbor distribution of energy levels $\Delta\epsilon$
shows a structure that can be fit with a Poisson distribution $P(S)=D^{-1}\exp({-S/D})$ with the local mean spacing $D=24.081$ cm$^{-1}$ (see, e.g., Ref. \citenum{Leitner}).
Such distributions are becoming of increasing interest due, e.g. to a recent
proposal \cite{Kauffman} regarding chaos and transport in biological processes.

The  upper  panel  of  Figure 1 shows torsional profiles of the
adiabatic   potential   energy   surfaces   (PES)  obtained  by
diagonalizing  the  Hamiltonian  matrix (\ref{model}) at $x=0$.
The  PES  profiles exhibit a conical intersection  at $\phi
\approx    \pi/2$   and   $x=0$,   clearly   visible   in   the
two-dimensional  plot  in the lower panel of Fig 1.
The  {\it  cis} isomer of retinal is localized in the potential
well  of  the lower diabatic PES ($n=0$) on the left-hand side.
Photoexcitation  by  incoherent  light  (represented by a green
arrow)  populates  a  number  of  stationary  eigenstates (grey
lines)  with mixed {\it cis}-{\it trans} character. The quantum
yield  is determined by projecting these eigenstates onto their
respective  \textit{cis}  and  \textit{trans}  regions  of configuration space as
discussed  above.  

Figure 2 shows the frequency-dependent quantum yield
$Y(\omega_i)$  given  by  Eq.  (\ref{QY2omega})  as  a  function  of
  excitation   energy   (measured   from   the  ground
vibrational  state  of  {\it  cis}-isomer). Here [from Eq.
(\ref{QY2omega})],  pure  {\it cis} states correspond to $Y = 0$
and  pure  {\it  trans}  states  correspond to $Y=1$; thus, the
magnitude  of  $Y(\omega_i)$  reflects  the  {\it  cis} or {\it
trans}   character  of  a  particular  eigenstate  with  energy
$\epsilon_i  =  \hbar\omega_i$.  As expected, all molecular eigenstates
that occur below the minimum energy
of   the   {\it   trans}   well  (11,000  cm$^{-1}$,  see  Fig. 2)  have  negligible  quantum  yields  due  to  the
absence  of  {\it  trans} eigenstates in this low energy range.
Above  the  11,000  cm$^{-1}$ threshold, the quantum yield is a
rapidly  varying  irregular function of $\omega_i$, reflecting
the  strong  mixing  between  the  {\it  cis}  and  {\it trans}
components  by  the full Hamiltonian. Such mixed {\it cis-trans}
eigenstates  are  conceptually	similar  to  the  long-lived
eigenstates  of  mixed  singlet-triplet  character  observed in
pentacene \cite{ZewailPNAS,ZewailSL}.

Interestingly, a closer inspection  of  Fig.  2,  as shown in the
figure  insert,  reveals the presence of purely {\it cis} (or {\it
trans})  eigenstates  with  $Y(\omega_i)  =  0$  (or 1), which are
qualitatively similar   to   the   electronically  localized  eigenstates  in
vibronically  coupled systems such as pyrazine\cite{Seideman}. 
The effect has recently been analyzed in the context of geometric phase-induced localization \cite{Artur,Loic}. 
However, such ``extreme states'' may well be a result of low dimensionality of the
model, as is indeed the case in pyrazine\cite{pyrazine}.

Table  I  lists  the values of the quantum yield computed using
the   stationary eigenstates   of   the   2D   model   (the  corresponding
$A_\text{2D}(\omega)$  is  shown  by  the  green  line  in Fig. 2)  \cite{Stock00,Stock03}.  Both  pre-averaged  and
post-averaged  results  ($Y_1  =  0.63$ and $Y_2 = 0.62$) agree
extremely well  with  the   previous
time-dependent    wavepacket    result\cite{Stock00} of $Y_\text{TD}=   0.63$.
 The  fact  that  the stationary quantum yields
agree  so  well with time-dependent calculations \cite{Stock00,Stock03}
makes clear how the quantum yield is directly
manifest in the stationary eigenstates. Moreover,  we also note  agreement with the measured value \cite{Mathies00} of $0.65\pm 0.01$. The agreement with experiment is a natural consequence of the two-state two-mode model \cite{Stock00} being parametrized to reproduce the measured quantum yield in time-dependent calculations.




In order to explore  the  effect of $A(\omega)$ on the
calculated  quantum  yields, we replaced the 2D model lineshape
function   in   Eqs.   (\ref{QY1})   and   (\ref{QY2})  by  the
experimentally   measured  absorption  profile  of  retinal  in
rhodopsin  \cite{Mathies00}  shown  by  the  red  line  in Fig. 2.  The result gives  $Y  =  0.43$, in worse agreement with
experiment  than the value obtained with $A_\text{2D}(\omega)$.
This confirms that the potential surfaces are optimized
for behavior in the domain shown in green in Fig. 2, but would
require further work to properly represent the \emph{cis/trans} branching
in other energy regions.


Mathies and co-workers \cite{Mathies00} also measured
$Y(\omega)$  observing a  decline  in  the
photoreaction efficiency below 500 nm. For comparison, we
calculate the wavelength-resolved stationary quantum yields by
dividing the  entire $\lambda$ interval into 10 nm-wide bins,
and  average  Eqs. (\ref{QY1}) and (\ref{QY2}) over
the eigenstates with energies falling into a particular bin. As
with  the overall quantum yield, two related, but not identical,
definitions are possible. In the first, one calculates
the   averaged   {\it   cis}   and   {\it   trans}  populations
\begin{equation}
\bar{P}_\alpha^{(n)} (\omega_i) = \sum_{j\in i\text{-th}\,\,\text{bin}} P_\alpha^{(n)}(\omega_j) A(\omega_j)
\end{equation}
where $\omega_i = 2\pi c/\lambda_i$ is the center frequency corresponding to $i$-th bin, and define:
\begin{equation}
Y_1(\omega_i) = \frac{ \bar{P}_\text{trans}^{(1)} (\omega_i) }{ \bar{P}_\text{trans}^{(1)} (\omega_i) + \bar{P}_\text{cis}^{(0)} (\omega_i) }
\end{equation}

As  an alternative, we directly average the frequency-dependent
quantum yield (\ref{QY2}) over the entire bin
\begin{equation}\label{Y2lambda}
Y_2(\lambda_i) = Y_2\left(\frac{2\pi c}{\omega_i}\right)= N_i^{-1} \sum_{j\in i\text{-th}\,\,\text{bin}} Y(\omega_j) A(\omega_j)
\end{equation}
where $N_i = \sum_{j\in i\text{-th}\,\,\text{bin}} A(\omega_i)$ is a normalization factor that serves to correct for the change in absorption intensity due to the varying $\lambda$.

Figure 3 compares the wavelength dependent
quantum yield with the experimental results in the 500-570 nm range
(Ref. \citenum{Mathies00}).
While  the  observed  quantum yield declines monotonically with
$\lambda$  and  stays  constant  below  $\lambda=500$  nm,  our
theoretical   values   oscillate   over  the  entire  range  of
$\lambda$.  Above 500 nm, the calculated quantum yields tend to
decline   with   $\lambda$,   in   qualitative  agreement  with
experiment (the only exception being the value of $Y_2$ at 540 nm). As expected from the above analysis (see Table I),
switching  lineshape  functions  has  a  dramatic effect on the
calculated   quantum  yields.  In  contrast  with  the  results
presented   in   Table   I,  however,  using  the  experimental
$A(\omega)$  improves  the  overall  agreement with experiment,
particularly at $\lambda > 500$ nm.

Also shown in the inset of Fig. 3 is the ``bare'' frequency-dependent quantum yield,
\begin{equation}\label{Ybare}
Y_\text{av}(\lambda_i) = N_i^{-1} \sum_{j\in i\text{-th}\,\,\text{bin}} Y(\omega_j).
\end{equation}
Here, the  absence of the spectral lineshape function [present in Eqs.~(\ref{QY1})  and  (\ref{QY2})]  reveals the
variation  in  {\it  cis}  / {\it trans} character of molecular
eigenstates  with  zero  oscillator  strengths,  which  do not
contribute  to the physical quantum yield (\ref{Y2lambda}). 
The $Y_\text{av}(\lambda)$ [inset of Fig. 3] is seen to generally decline  with  increasing  $\lambda$. This
behavior  reflects the appearance of {\it trans}
states  at energies  above  11,000  cm$^{-1}$,  and  that their density in this
region  is  smaller  than  that  of  {\it cis} states (see Fig. 2).  Thus,  the  {\it trans} character of molecular
eigenstates   can  be  expected  to  decrease  with  decreasing
$\omega_i$, as is implicit in the experimental results. While the downward trend in the wavelength dependence of $Y_\text{av}$ (inset of Fig. 3) is somewhat more pronounced than that observed for $Y_1$ and $Y_2$ (Fig. 3), we note that the former cannot be directly compared with experiment. 
This is because the bare quantum yield [Eq. (\ref{Ybare})] does not
include the variation of the spectral lineshape function 
$A(\omega)$ with $\omega$.

\section{Effects of relaxation on quantum yield}

Our computed time-independent definitions of the quantum yield 
[Eq. (\ref{QY1}) and (\ref{QY2})] assume that the eigenstate populations $\rho_{ii}$ do not depend on time. In reality, however, the populated eigenstates undergo relaxation due to the interaction with the environment. To elucidate the effect of the relaxation on the quantum yield, we calculated the transition rates $W_{j\leftarrow i}$ between eigenstates $|i\rangle$ and $|j\rangle$ using Fermi's Golden Rule \cite{Blum}
\begin{equation}\label{relax1}
W_{j\leftarrow i} =\sum_{\gamma=\phi,x} |\langle i | \hat{Q}_\gamma | j \rangle|^2 (1+N(|\omega_{ji}|)) J_\gamma(|\omega_{ji}|) \quad (\omega_{ji}<0)
\end{equation}
and
\begin{equation}\label{relax2}
 W_{j\leftarrow i}=\sum_{\gamma=\phi,x} |\langle i | \hat{Q}_\gamma | j \rangle|^2 N(\omega_{ji}) J_\gamma(\omega_{ji}) \quad (\omega_{ji}>0),
\end{equation}
where $N(\omega) = [\exp (\hbar\omega/kT) -1]^{-1}$ is the Bose distribution at temperature $T=300$ K, $k$ is the Boltzmann's constant, and $J_\gamma(\omega)=\eta_\gamma e^{-\omega_\gamma/\omega_{c\gamma}}$ is the spectral density of the bath modes representing low-frequency, non-reactive vibrational modes \cite{Stock00a} coupled to degree of freedom $\gamma=\phi,\,x$. Following Stock and co-workers \cite{Stock03}, we adopt an Ohmic spectral density $J(\omega)=\eta_\gamma e^{-\omega/\omega_{c\gamma}}$ with $\eta_\phi = 0.15$, $\omega_{c\phi}=0.08$~eV, $\eta_x = 0.1$, and $\omega_{cx}=0.19$ eV. The bath operators $\hat{Q}_\gamma$ in Eqs.~(\ref{relax1})-(\ref{relax2}) are given by \cite{Stock05} $\hat{Q}_\phi = (1-\cos\phi) |\psi_1\rangle \langle \psi_1|$ and $\hat{Q}_x = x|\psi_1\rangle\langle \psi_1|$.

Figure 4(a) shows the time dependence of the expectation values $P^{(n)}_\text{cis}(t)$ and $P^{(n)}_\text{trans}(t)$ in Eq. (\ref{populations}) obtained  by propagating the rate equations parametrized by the transition rates given by Eqs. (\ref{relax1})-(\ref{relax2})  with the initial condition $\rho_{ij}(t=0)=\delta_{ij}A(\omega_i)$, {\it i.e.} assuming fully incoherent excitation with natural light. 
A substantial fraction of population at $t=0$  resides in the excited diabatic electronic states [$P^{(1)}_\text{cis}+P^{(1)}_\text{trans}=0.69$]. The interaction with low-frequency  bath modes leads to dissipation of the electronic and vibrational energy, manifested in the decay of the excited-state populations $P^{(1)}_\text{cis}(t)$ and $P^{(0)}_\text{trans}(t)$. The decay is accompanied by a growth of the ground-state populations $P^{(0)}_\text{cis}(t)$ and $P^{(1)}_\text{trans}(t)$ shown by the solid lines in Fig. 4(a). 

Figure  4(b)  plots  the  time  dependence of the quantum yield
given  by  Eq.  (\ref{Y}).  Remarkably, the quantum yield shows
only   a   weak  time  dependence,  with  deviations  from  the
asymptotic  value  of 0.62 not exceeding 3\%  
over the time interval studied (0 -- 3 ps). Thus,
while  the  individual  {\it  cis}  and {\it trans}-populations
evolve  in  time,  the  value  of the quantum yield, defined as
their  ratio via Eq. (\ref{QY1}), remains constant. 

Analysis of the system-bath coupling matrix elements [Eq. (\ref{relax1})] shows that the matrix elements involving the torsional degree of freedom $\phi$ are small compared to those of the coupling mode $x$. We can therefore expect that relaxation of the initial ``bright'' eigenstates populated by incoherent FC excitation (see below and Fig. 5) is driven by the interaction of the coupling mode with the bath oscillators.  Indeed, as shown in Fig. 4(b), neglecting the $\phi$-component of the system-bath coupling 
(dashed curve) has little effect on the time variation of the quantum yield.  The dominant role played by the coupling mode in the relaxation process is at the heart of our arguments presented below that analyze the lack of time dependence of the quantum yield.



In order to gain insight into relaxation dynamics, we plot in Fig. 5 the transient linear absorption spectrum of the two-state, two-mode model [e.g. the populations $\rho_{ii}(t)$] during the various stages of the relaxation process. At time zero, an incoherent mixture of ``bright'' eigenstates is assumed to follow FC excitation from the ground state. The  three dominant ``bright'' eigenstates that account for over 40\% of all $t=0$ excited-state population are $|512\rangle$, $|507\rangle$, and $|508\rangle$, and are focused upon below. As time proceeds, the eigenstates begin to relax due to the interaction with the phonon bath. At $t=100$~fs, the population is seen to be 
spread over three lower-lying manifolds of states, which are separated  from the initial ``bright'' manifold by a constant energy gap. The dominant eigenstates populated in the decay of the bright state $|512\rangle$, are shown in Fig.~6. We observe that during the course of relaxation, the population ``branches out'' into different final eigenstates until it finally settles in a steady state characterized by a stationary eigenstate distribution shown in the lowermost panel of Fig. 5. The steady state defines the asymptotic ($t\to \infty$) limit of the quantum yield, and is thus of particular importance to the theoretical description. 

According to Fermi's Golden Rule (\ref{relax1}), the relaxation rates are determined by (a) the magnitude of the system-bath coupling matrix element, and (b) the value of the spectral density at the transition frequency $J(\omega_{ij})$. 
The existence of multiple intermediate relaxation stages in Fig. 6 is a consequence of a fixed cutoff frequency of the bath $\omega_{cx}=0.19$ eV. Because the Ohmic spectral density function peaks at $\omega = \omega_{cx}$ and decays quickly away from the maximum \cite{MayKuhn}, relaxation to the eigenstates separated from the initial state by the energy window $\hbar\omega_{cx}\pm \hbar\Delta\omega$ is allowed (provided that the corresponding coupling matrix elements are non-zero), whereas relaxation to the eigenstates outside the energy window cannot occur via single phonon-emission. However, such lower-lying eigenstates are populated via multiple phonon emission (or relaxation cascade), as illustrated in Fig. 6.

As for the quantum yield, relaxation influences the quantum yield through the time dependence of the populations $\rho_{ii}(t)$ which changes the eigenstates that contribute to the evolving quantity:
\begin{equation}\label{QY1t}
Y_1 (t) = \frac{ \sum_i \rho_{ii}(t) \langle i |\hat{P}^{(1)}_\text{trans}|i\rangle } { \sum_i \rho_{ii}(t) \langle i|\hat{P}^{(0)}_\text{cis}|i\rangle + \sum_{i} \rho_{ii}(t) \langle i|\hat{P}^{(1)}_\text{trans}|i\rangle }
\end{equation}
Therefore, in order to understand the time dependence of the quantum yield, it is necessary to follow both the time evolution of the populations $\rho_{ii}(t)$ and their localization properties, as manifest in the matrix elements of $\hat{P}^{(0)}_\text{cis}$ and $\hat{P}^{(1)}_\text{trans}$. 
To examine the localization, we define the probability density $P_i^{(n)}(\phi,x)$ for eigenstate $|i\rangle$ projected onto the diabatic electronic state $n=0,1$
\begin{equation}\label{probden}
P^{(n)}_i(\phi,x) = \int d\mathbf{r}_e |\psi_i^{(n)}(\phi, x; \mathbf{r}_{e})|^2,
\end{equation}
where $\psi_i^{(n)}(\phi, x; \mathbf{r}_{e})$ are the diabatic components of the eigenstate $|i\rangle$ described by the wavefunction 
\begin{equation}\label{expansion1}
\psi_i(\phi, x; \mathbf{r}_{e}) = \sum_{n=0,1} \psi_i^{(n)}(\phi, x; \mathbf{r}_{e}) = \sum_{n=0,1} \sum_{\mu,\nu} C_{n\mu\nu} \Phi^{(n)}_\mu (\phi) X^{(n)}_{\nu}(x) \psi_n(\mathbf{r}_e),
\end{equation}
where the direct-product basis $\Phi^{(n)}_\mu (\phi) X^{(n)}_{\nu}(x)$ is chosen in such a way as to diagonalize the torsional vibronic coupling Hamiltonian (\ref{model})  at $\lambda=0$ (no vibronic coupling). In this latter case, the problem reduces to that of two non-interacting electronic states, and the eigenstates (\ref{expansion1}) factorize into  products of basis functions that depend on the $\phi$ and $x$ coordinates 
\begin{equation}\label{zeroth_order}
\psi_i(\phi, x; \mathbf{r}_{e}) = \Phi^{(n)}_\mu (\phi) X^{(n)}_{\nu}(x) \psi_n(\mathbf{r}_e),
\end{equation}
This expression for the system eigenstates provides a reasonable zeroth-order approximation to the true eigenstates of the vibronic coupling Hamiltonian (\ref{model}) in the energy region below the conical intersection ($\epsilon < 14,000$ cm$^{-1}$) \cite{Tscherbul}. Here, the low-lying eigenstates of the full vibronic Hamiltonian are localized in their own potential wells \cite{Artur,Loic} (as clearly observed in Fig. 2). However, as the energy increases above the conical intersection, the eigenstates become strongly mixed by the vibronic coupling, and delocalize over the whole configuration space.

An example of such behavior is shown in Fig. 7, which shows the probability density of the bright eigenstate $|512\rangle$ together with that of the ground state $|1\rangle$ and the eigenstates populated in the first few relaxation stages (green arrows in Fig. 6). The bright state has a large probability amplitude in the {\it cis}-region of the first excited diabatic state, which maximizes its overlap with the ground {\it cis}-state, to enhance excitation. 
The lower-lying states  $|419\rangle$ and $|423\rangle$ start to experience the repulsive wall of the $n=1$ diabatic potential in the {\it cis}-region of configuration space (see Fig. 1). As a result, the population is completely transferred away from the repulsive region to the ground {\it cis} and excited {\it trans} regions. There is no apparent preference for either {\it cis} or {\it trans} regions during these initial relaxation stages due to the strong mixing of all degree of freedom: As shown in Fig. 4(a), both $P^{(0)}_\text{cis}$ and $P^{(1)}_\text{trans}$ increase monotonically with time.

After being transferred one step below below the initial state, the population continues to relax to lower-lying eigenstates while preserving its profile along the reaction coordinate. This 
can be attributed to the particular form of the system-bath coupling, which primarily depends on the $x$ coordinate and hence does not strongly alter the $\phi$-profiles of the  eigenstates coupled by the bath. In particular, we verified that  the amount of {\it cis} and {\it trans} character of most of the states involved in the relaxation process is approximately constant. Exceptions do occur because of the strong vibronic coupling, which may occasionally change the $\phi$ distribution of some eigenstates due to the coupling mediated by the $x$ coordinate. However, this situation is an exception rather than the rule:  As shown in Fig. 8, only two out of $\sim$20 eigenstates involved in the first stage of relaxation ($|421\rangle$ and $|415\rangle$) are dramatically different in their {\it trans}-character (as quantified by the matrix element $\langle i| P^{(1)}_\text{trans}|i\rangle$) from the other states.   As a result, the quantum yield varies only weakly with time.


A dramatic change in the relaxation mechanism occurs when the eigenstate energy falls below that of the conical intersection. Figure 9 shows that the eigenstates 
to which relaxation then occurs become localized in their  corresponding {\it cis-} and {\it trans}- wells. The population of the initial (delocalized) eigenstate $|346\rangle$ relaxes to two eigenstates, one of which  ($|279\rangle$) is strongly localized in the {\it trans}-well and the other ($|281\rangle$) has more probability amplitude in the {\it cis}-well (see Supplementary Material \cite{SI}). As the population gets partitioned between the two eigenstates, the quantum yield remains unaltered again due to the $\phi$-independent nature of the system-bath coupling, which tends to conserve the angular probability density of all eigenstates involved.

Finally, we consider the final stages of the relaxation process depicted in Fig. 9. These represent bath-induced transitions between localized eigenstates, and can be understood by noting that the eigenstates are given to zeroth order by Eq. (\ref{zeroth_order}). For simplicity, let us assume that the value of the quantum yield before relaxation begins is determined by a single eigenstate $\alpha$. After relaxation is over, the population is transferred to the eigenstate $|\alpha'\rangle$.  Taking the matrix elements of $\hat{P}^{(1)}_\text{trans}$ and using Eq. (\ref{zeroth_order}), we find
\begin{equation}\label{loc1}
\langle \alpha |\hat{P}^{(1)}_\text{trans}|\alpha\rangle = \langle \Phi_\mu^{(1)}(\phi) X_\nu^{(1)}(x) | \hat{P}^{(1)}_\text{trans} | \Phi_\mu^{(1)}(\phi) X_\nu^{(1)}(x)\rangle = \langle \Phi_\mu^{(1)}(\phi) | \hat{P}^{(1)}_\text{trans} | \Phi_\mu^{(1)}(\phi) \rangle 
\end{equation}
 because $\hat{P}^{(1)}_\text{trans}$ does not depend on $x$, and similarly
\begin{equation}\label{loc2}
\langle \alpha' |\hat{P}^{(1)}_\text{trans}|\alpha'\rangle = \langle \Phi_{\mu'}^{(1)}(\phi) | \hat{P}^{(1)}_\text{trans} | \Phi_{\mu'}^{(1)}(\phi) \rangle.
\end{equation}
But the system-bath coupling operator does not depend on $\phi$, and hence cannot change the number of quanta in the torsional mode $\mu$, so the right-hand sides of Eqs. (\ref{loc1}) and (\ref{loc2}) are equal. 

This result suggests that in the low-energy regime of localized eigenstates, the quantum yield of {\it any} process that depends on a single reaction coordinate should be time-independent, provided that the system-bath coupling does not depend on that coordinate. This is clearly illustrated in Fig.~9: the localized state $|279\rangle$ has 3 peaks along the $x$ coordinate, corresponding to 2 quanta in the $x$-mode ($\nu=2$). The system-bath coupling changes $\nu$ from 2 to 1 and then from 1 to 0, while leaving the $\phi$ distribution unchanged (see also Figs. 1 and 2 of Supplementary Material \cite{SI}). The final eigenstate $|171\rangle$ is vibrationally ``cold'' with respect to the coupling mode, but remains highly excited along the torsional coordinate.

A few closing remarks are in order concerning the asymptotic ($t\to \infty$) value of the quantum yield that corresponds to the population distribution shown in the lowermost panel in Fig. 6. The steady state is obtained by propagating the rate equations of motion for the populations, which is a numerically efficient procedure that scales quadratically with the number of eigenstates \cite{Blum}.  The nature of the steady state is determined by the form of the system-bath coupling, and the initial conditions. We emphasize that this  steady-state solution does \emph{not}
have the form of the Boltzmann distribution that would be normally expected of Pauli-type rate equations in the 
limit $t\to \infty$ \cite{Reichman}. Rather, 
in accord with Ref. \citenum{Stock02}, our steady-state distribution corresponds to a metastable state,  which will eventually tunnel to the {\it cis}-well (at least if the well is one-dimensional). The large barrier height in the two-state two-mode model makes the tunneling timescale extremely long compared to any other timescale of interest in this system.
 As a possible direction of future research, it would be interesting to explore whether such metastable steady states can be obtained without propagating the dynamical equations of motion.
  If so, the determination of the quantum yield from the stationary eigenstates becomes a straightforward task via Eq. (\ref{QY1}) with the matrix elements $\rho_{ii}$ replaced by their steady-state values.

\section{Summary and Discussion}
\label{summary}

We have considered a time-independent approach to the quantum yield of
\textit{cis-trans} photoisomerization, here applied to model retinal in rhodopsin. 
The need for this approach arises due to the recognition that natural processes
take place in incoherent light (e.g., sunlight with a coherence time of
1.32 fs) and environmental decoherence, which produce mixtures of  stationary
Hamiltonian eigenstates.  Here we have recast one of the 
standard time-dependent definitions of  the quantum yield 
in terms  of  time-independent  quantities,  the  eigenstates of
the Hamiltonian and associated dipole transition matrix elements.
The
quantum yield is then shown to be a direct reflection of the {\it cis} vs.
{\it trans} character of the individual stationary eigenstates 
of the system
and the associated dipole transition matrix elements from
the ground electronic state. Further,
applied to a model of retinal, this approach gives excellent results
for the quantum yield, fully in agreement with experiment.
Interestingly, relaxation from the initially prepared stationary 
mixture does not alter the quantum yield, a consequence of both
the {\it cis/trans} partitioning of the stationary states and the system-bath
coupling in this well established minimalist model of retinal isomerization.

Ideally, we should consider a full treatment of all modes of
retinal in a proper rhodopsin environment, define the quantum
yield and other observable properties, fit the retinal
potential parameters to experiment, and compare time-dependent
computational results to the stationary state results associated with
incoherent light excitation. However, such and
extensive computational study is not
required to support the main result of this paper, i.e.,
that the stationary eigenstates of the system Hamiltonian provide an
alternative and important way to understand features affecting the
quantum yield in incoherent light. Rather, we adopt the basic
two-dimensional model of Hahn and Stock \cite{Stock00,Stock00a}. Restricting attention
to this model necessitates, by requirements of consistency, that
if we adopt their two-dimensional potentials and associated
system parameter fits to experimental data, that we also must
maintain their definition of quantum yield, with the associated
neglect of $P^{(1)}_\text{cis}(t)$ and $P^{(0)}_\text{trans}(t)$. Alternative definitions of the
quantum yield  would have resulted in different values of the
system parameter fits and different stationary
eigenstates of the Hamiltonian. Hence, the successful computational results
demonstrated here do motivate a more
extensive calculation of retinal Hamiltonian eigenstates for a
retinal model including all degrees of freedom\cite{Arango}. Such work is in
progress.


\textbf{Acknowledgments}.  This work was supported by the Natural Sciences and Engineering Research Council of Canada, and by the US Air Force Office of Scientific Research under contract numbers FA9550-10-1-0260
and FA9550-13-1-0005.

\newpage

Table I.
Calculated quantum yields for retinal photoisomerization. Both pre-averaged ($Y_1$) and post-averaged ($Y_2$) results are shown. $A_\text{2D}(\omega)$ -- normalized lineshape function calculated within the 2D model; $A_\text{exp}(\omega)$ -- normalized lineshape function based on the measured absorption spectrum of retinal in rhodopsin \cite{Mathies00}. The time-dependent wavepacket result from Ref. \citenum{Stock00} is given in the last column.

\centering
\vspace{0.3cm}
\begin{tabular}{ccccc}
\hline
\hline
 Quantum yield $\quad$ & $\quad$ $A_\text{2D}(\omega)$ $\quad$ & $\quad$ $A_\text{exp}(\omega)$ $\quad$ & Experiment \cite{Mathies00} & Time-dependent \cite{Stock00} \\
\hline
\hline
$Y_1$ & 0.62 & 0.43 & $0.65\pm 0.01$ & 0.63 \\
$Y_2$ & 0.63 & 0.45 & & \\
\hline
\hline
\end{tabular}\begin{flushleft}
\end{flushleft}




\newpage

\begin{figure}[!t]
	\centering
	\includegraphics[width=1.03\textwidth, trim = 0 0 0 0]{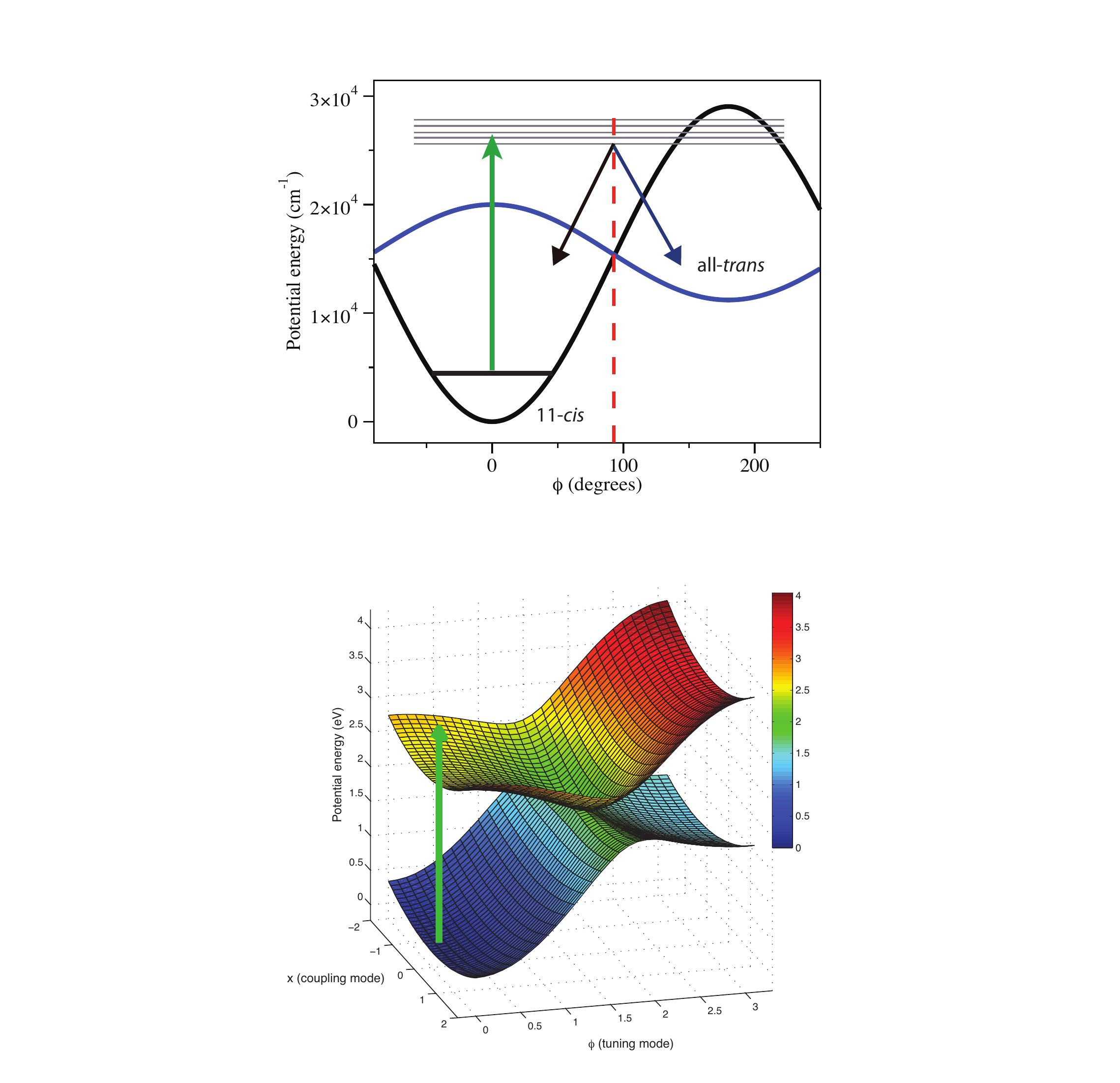}
	\renewcommand{\figurename}{Fig.}
	\caption*{}\label{fig:pes}
\end{figure}

\flushleft
\small Fig. 1: (Upper panel) Ground and first excited diabatic potential profiles along the reaction coordinate $\phi$ for retinal photoisomerization. The green upward arrow illustrates laser excitation, the downward arrows illustrate the partitioning of the eigenstates into \textit{cis} and \textit{trans} (by the projection operators, see text). The {\it cis} and {\it trans} regions of configuration space are separated by the red dashed line. (Lower panel) Adiabatic PESs for retinal as functions of the torsional coordinate $\phi$ and the coupling mode $x$ orthogonal to it. The PESs are obtained by diagonalizing the Hamiltonian in Eq. (\ref{model}) (without the kinetic energy term).  \rm

\newpage

\begin{figure}[t]
	\centering
	\includegraphics[width=0.7\textwidth, trim = 0 0 0 0]{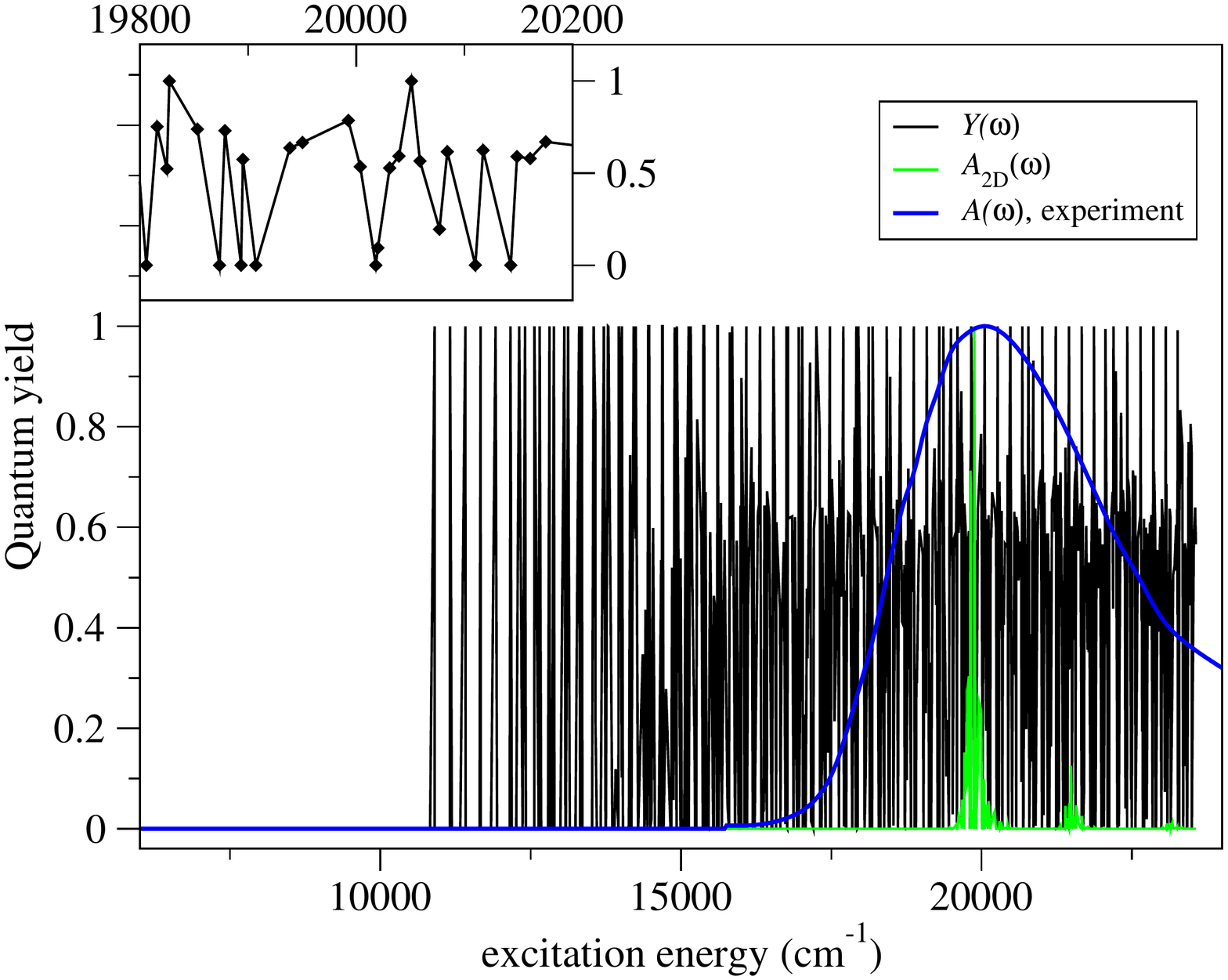}
	\renewcommand{\figurename}{Fig.}
	\caption*{}\label{fig:qy}
\end{figure}

\flushleft
\small Fig. 2: Frequency dependence of the stationary quantum yield. Superimposed on the plot are the linear absorption spectrum of the 2D model (green) and the experimental absorption spectrum spectrum of retinal in rhodopsin adapted from Fig. 2 of Ref. 23, both normalized to unity at their respective maxima. The inset shows an expanded view of the frequency-dependent quantum yield in the region of maximum absorption at $\omega\sim 20,000$ cm$^{-1}$ ($\lambda = 500$ nm) \rm

\newpage

\begin{figure}[t]
	\centering
	\includegraphics[width=0.7\textwidth, trim = 0 0 0 0]{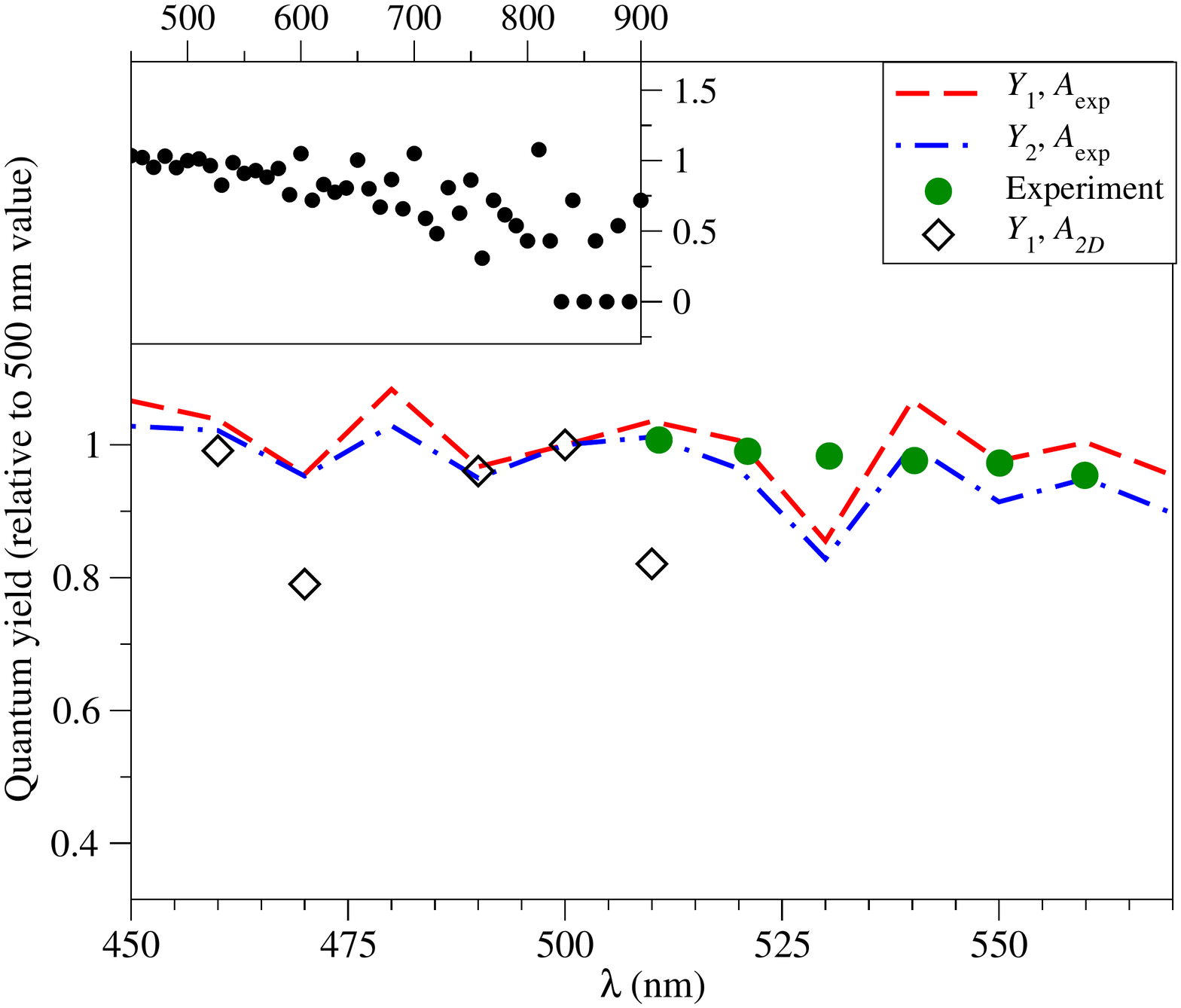}
	\renewcommand{\figurename}{Fig.}
	\caption*{} \label{fig:exp}
\end{figure}

\flushleft
\small Fig. 3: Wavelength dependence of the stationary quantum yield. Diamonds -- results for $A_\text{2D}(\omega)$ calculated from the two-state two-mode model; dashed (dash-dotted) lines -- results obtained for $Y_1$ ($Y_2$) and the experimentally measured $A(\omega)$ \cite{Mathies00}, circles -- experiment. The error bars are smaller than the size of the circles. The inset shows the bare frequency-dependent quantum yield (\ref{Ybare}) calculated {\it without} the spectral lineshape function $A(\omega)$. The experimental quantum yield stays constant below $\lambda = 500$ nm. The results for $A_\text{2D}(\omega)$ are shown only in those spectral regions where $A_\text{2D}(\omega)$ does not vanish (see Fig. 2).

\newpage

\begin{figure}[t]
	\centering
	\includegraphics[width=0.7\textwidth, trim = 0 0 0 0]{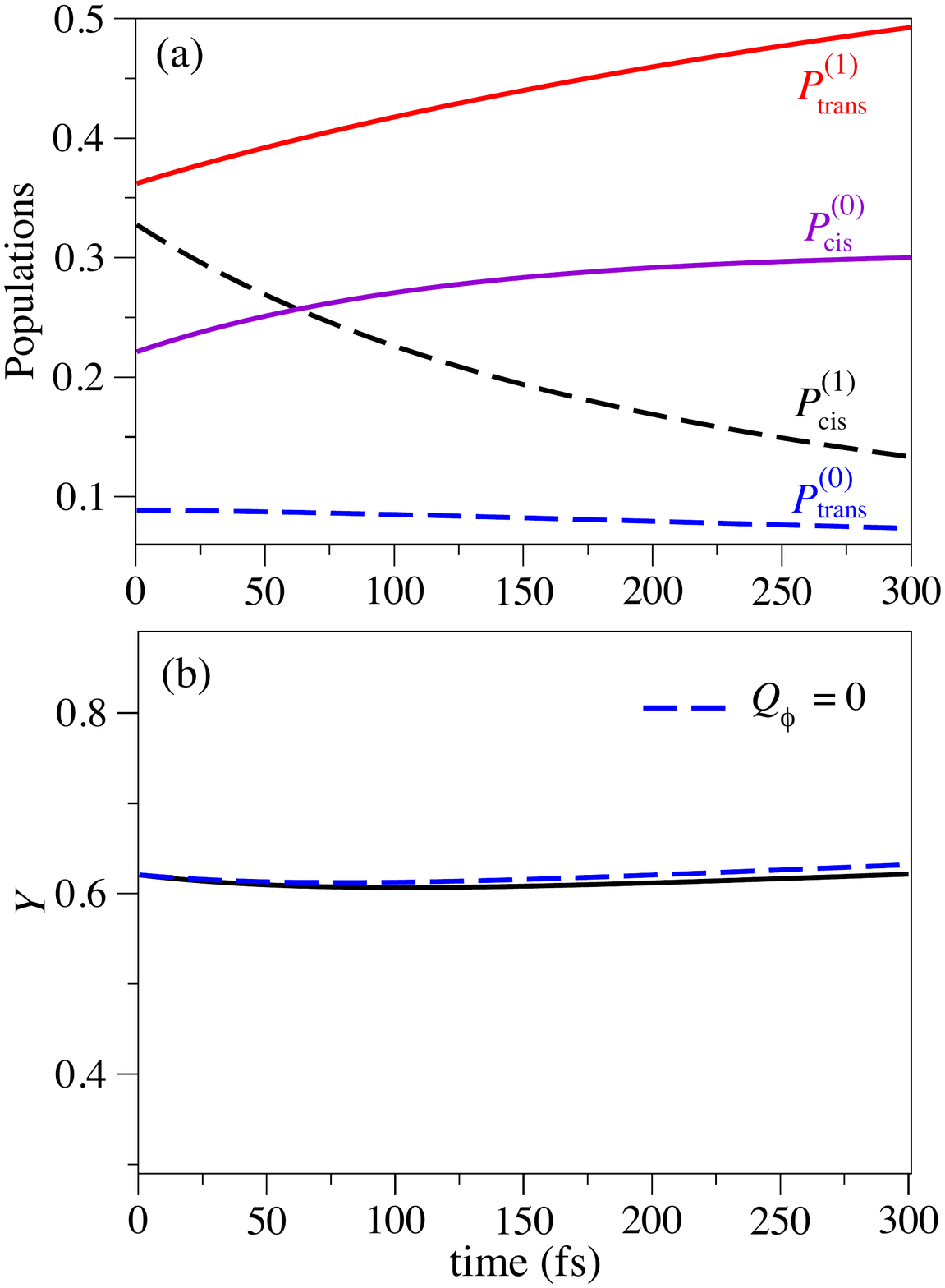}
	\renewcommand{\figurename}{Fig.}
	\caption*{} \label{fig:exp}
\end{figure}

\flushleft
\small Fig. 4: (a) Time dependence of {\it cis} and {\it trans} state populations (see Eq. \ref{populations}). The excited-state populations $P^{(1)}_\text{cis}$ and $P^{(0)}_\text{trans}$ (dashed lines) decay in time due to the interaction with the environment (see text). (b) Time dependence of the quantum yield given by Eq. (\ref{Y}). These results are obtained by solving the equations of motion for the diagonal elements of the density matrix parametrized by the transition rates given by Eqs.~(\ref{relax1})-(\ref{relax2}). The quantum yield remains constant (within 3\% of the asymptotic value of 0.62) over the time interval $0<t<3000$ fs.

\newpage
\begin{figure}[t]
	\centering
	\includegraphics[width=0.7\textwidth, trim = 0 0 0 0]{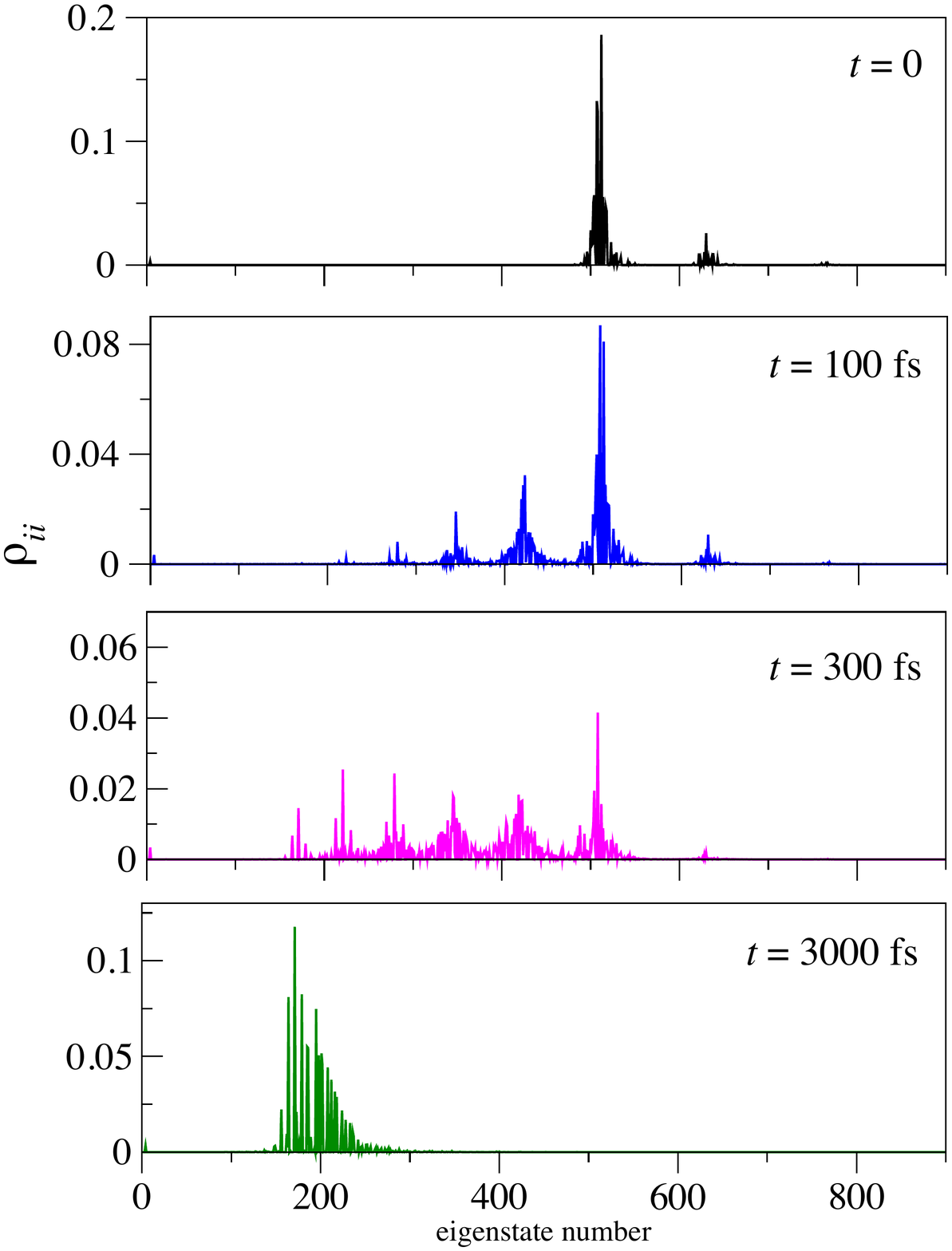}
	\renewcommand{\figurename}{Fig.}
	\caption*{} \label{fig:exp}
\end{figure}
\flushleft
\small Fig. 5: Snapshots of eigenstate populations $\rho_{ii}(t)$. At $t=0$, the bright eigenstates (mostly 512, 507, and 508) are populated by fully incoherent, impulsive FC excitation (see Eq. 6). At later times, interaction with the bath causes the population of the bright states to decay through several cascades (middle panels). The resulting steady-state eigenstate distribution is plotted in the lowermost panel.

\newpage
\begin{figure}[t]
	\centering
	\includegraphics[width=0.7\textwidth, trim = 0 0 0 0]{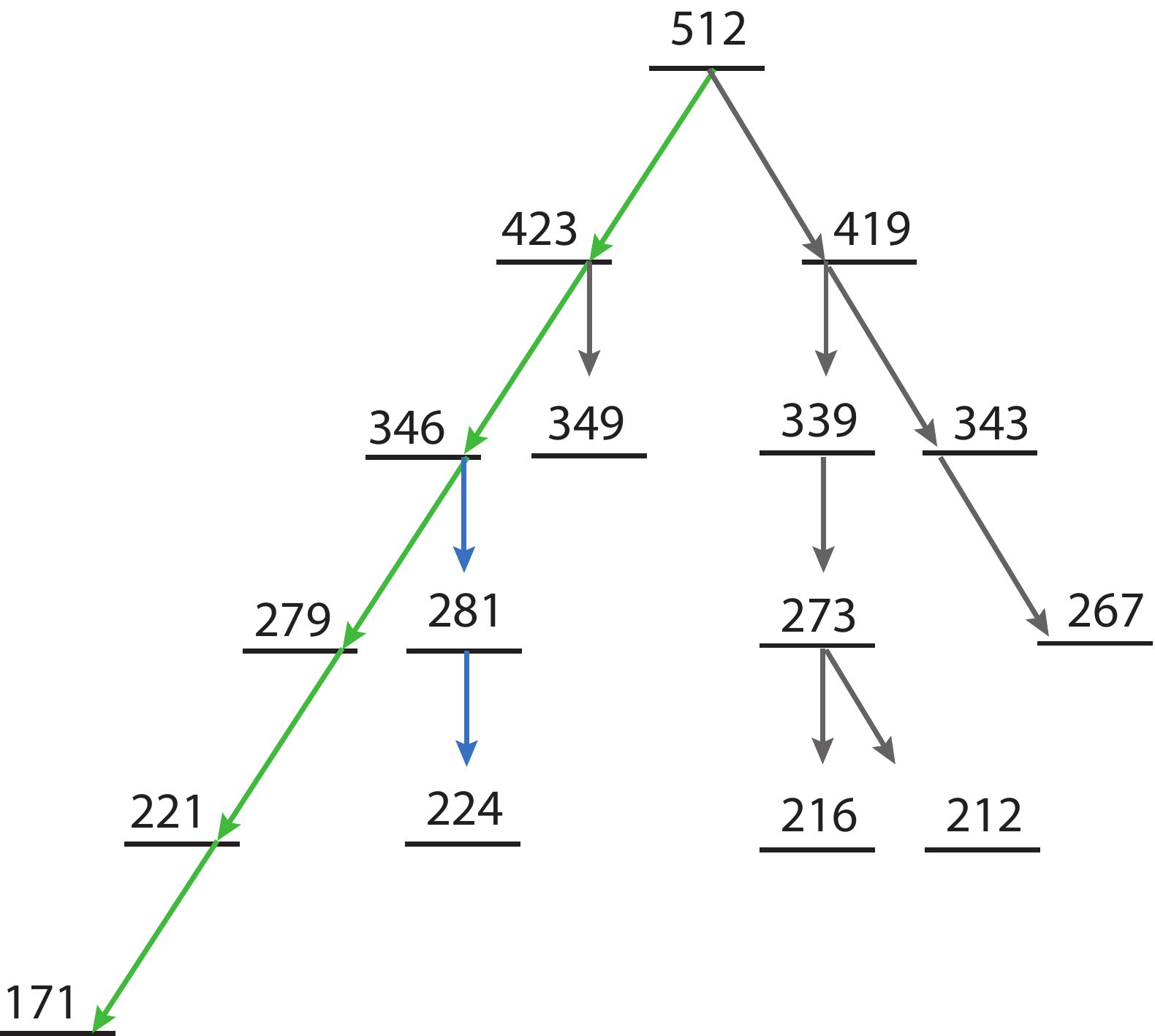}
	\renewcommand{\figurename}{Fig.}
	\caption*{} \label{fig:exp}
\end{figure}
\flushleft
\small Fig. 6: The network of dominant relaxation pathways starting from the dominant bright state $|512\rangle$ (the highest peak in the upper panel of Fig. 5). Green arrows show the most efficient pathway leading to the $|171\rangle$ eigenstate (the highest peak in the lower panel of Fig. 6). See text for  details.

\newpage
\begin{figure}[t]
	\centering
	\includegraphics[width=0.7\textwidth, trim = 0 0 0 0]{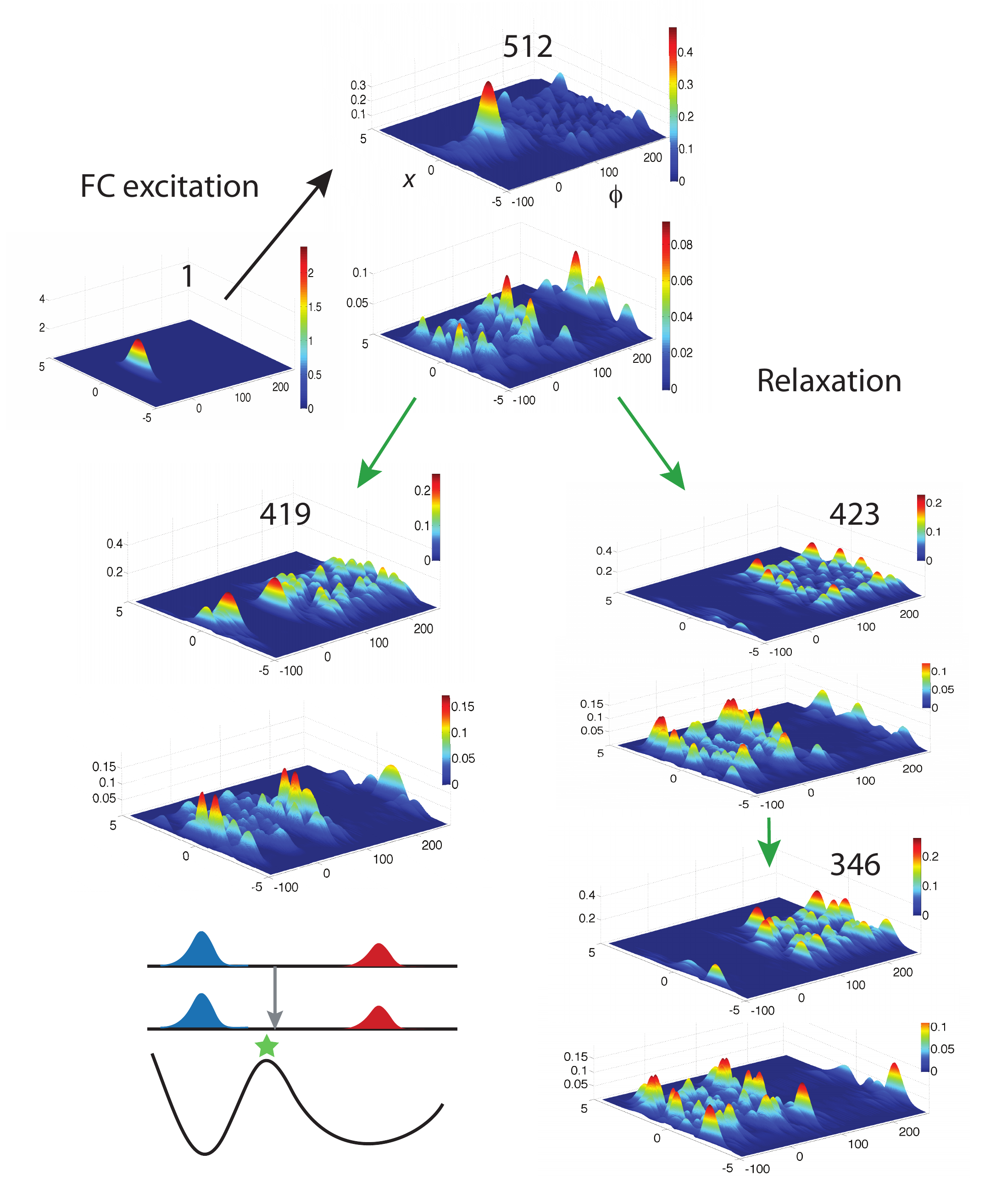}
	\renewcommand{\figurename}{Fig.}
	\caption*{} \label{fig:exp}
\end{figure}
\flushleft
\small Fig. 7: Probability density (Eq. \ref{probden}) for the ground ($n=0$, upper panels) and the first excited ($n=1$, lower panels) diabatic electronic states as a function of the torsional coordinate $\phi$ and the coupling mode $x$. Only the eigenstates above the conical intersection are shown. Note the similarity of the $\phi$-profiles of the eigenstates involved in the relaxation process, which implies no change in the quantum yield. The lower left figure illustrates that {\it cis/trans} partitioning of delocalized eigenstates above the conical intersection  (marked by the star) does not change qualitatively during relaxation. Note changes in color scale from panel to panel.

\newpage
\begin{figure}[t]
	\centering
	\includegraphics[width=0.7\textwidth, trim = 0 0 0 0]{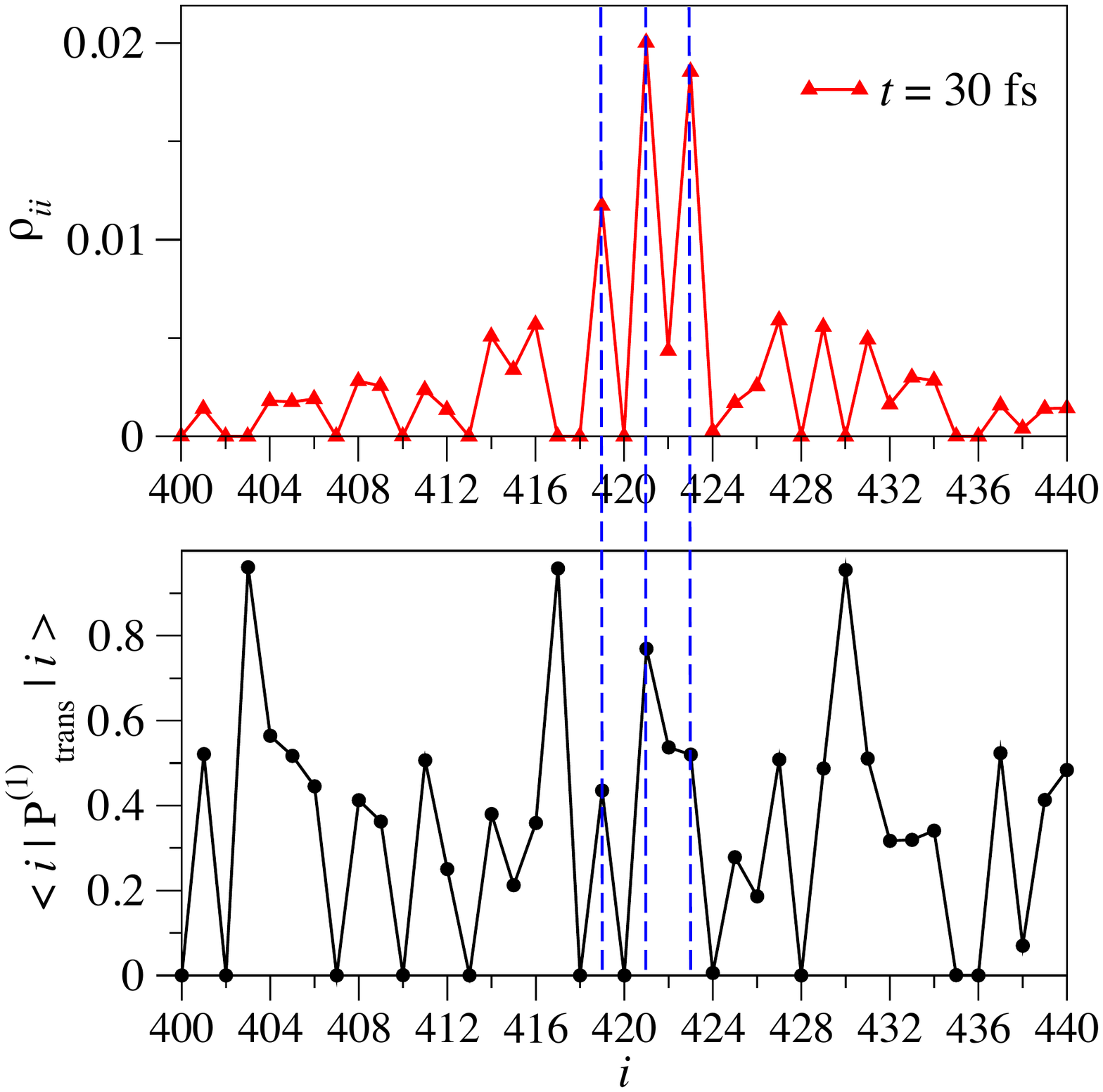}
	\renewcommand{\figurename}{Fig.}
	\caption*{} \label{fig:exp}
\end{figure}
\flushleft
\small Fig. 8: Localization properties of the eigenstates in the first relaxation cascade. (Upper panel): the populations of the first-cascade eigenstates at $t=30$  fs after the initial excitation. (Lower panel): the diagonal matrix element of the projection operator $\hat{P}^{(1)}_\text{trans}$ quantifying the amount of population in the {\it trans}-region of configuration space.   Note that all eigenstates significantly populated in the first stage of relaxation have similar localization properties with $\langle i|P^{(1)}_\text{trans}|i\rangle = 0.36 - 0.56$. Notable exceptions include states $|421\rangle$  and  $|415\rangle$ with $\langle 421|P^{(1)}_\text{trans}|421\rangle = 0.76$ and $\langle 415|P^{(1)}_\text{trans}|415\rangle = 0.21$.

\newpage
\begin{figure}[t]
	\centering
	\includegraphics[width=0.7\textwidth, trim = 0 0 0 0]{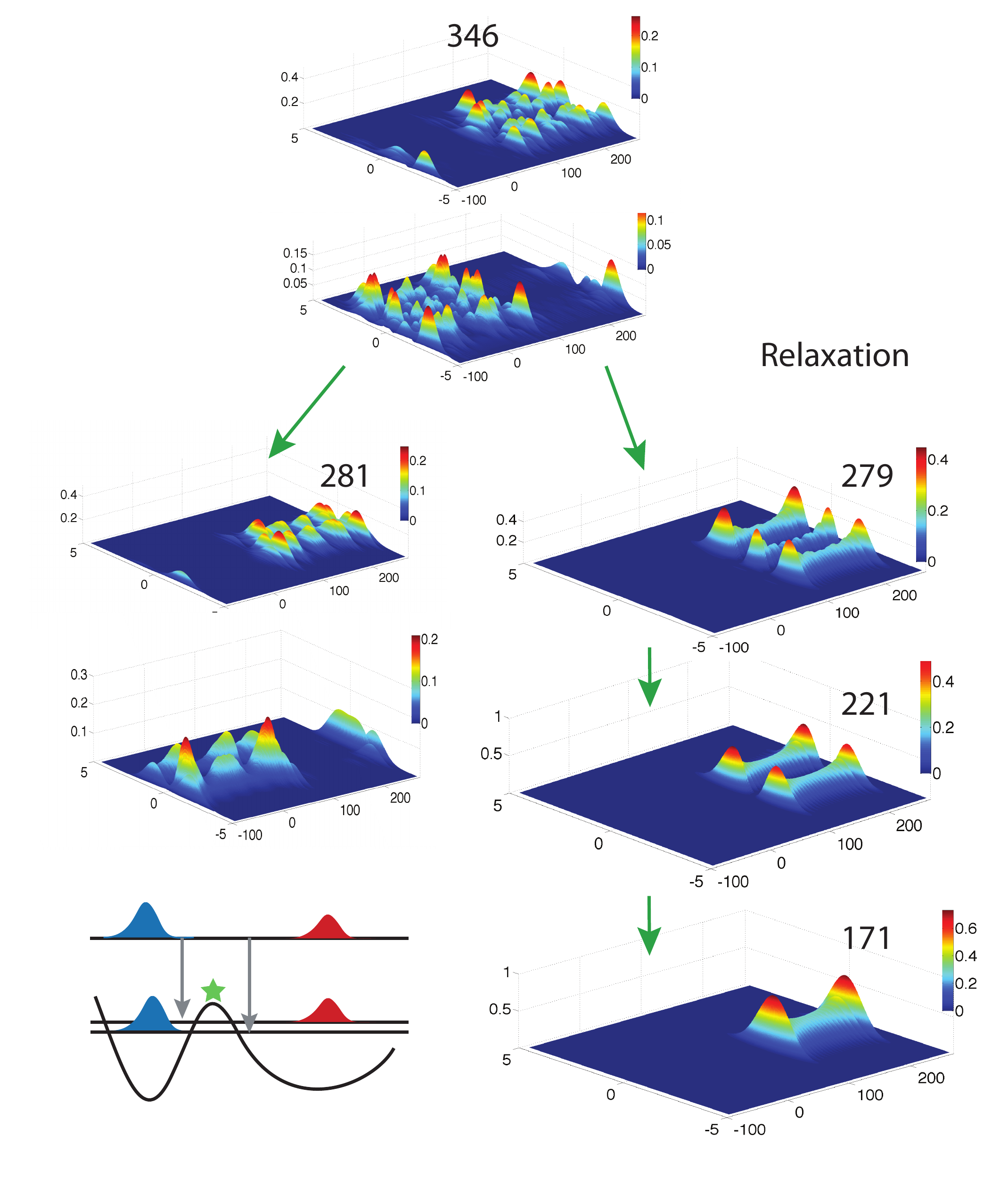}
	\renewcommand{\figurename}{Fig.}
	\caption*{} \label{fig:exp}
\end{figure}
\flushleft
\small Fig. 9 Same as in Fig. 7 but for the eigenstates in the vicinity of and below the conical intersection. Note the different localization properties of the eigenstates involved in the relaxation process. The lower left figure illustrates the transition from delocalized to localized eigenstates below the conical intersection (marked by the star).
Note changes in color scale from panel to panel.


\end{document}